\documentclass[12pt]{article}
\usepackage{epsfig}

\newcommand\pubnumber{CERN-TH/2001-48\\
MPI-PhT/2001-01}
\newcommand\hepnumber{hep-ph/0102292}
\newcommand\pubdate{February 2001}

\def\csumb{
Theory Division, CERN,\\
   CH-1211 Geneva 23, Switzerland
}
\def\support{\footnote{Permanent address: 
Max-Planck-Institut f\"ur Physik,
Werner-Heisenberg-Institut,
F\"ohringer Ring 6, D-80805 M\"unchen, Germany.}} 

\def\Title#1{\begin{center} {\Large\bf #1 } \end{center}}
\def\Author#1{\begin{center}{ \sc #1} \end{center}}
\def\Address#1{\begin{center}{ \it #1} \end{center}}

\newcommand\pubblock{\rightline{\begin{tabular}{l} \pubnumber\\
         \pubdate\\ \hepnumber \end{tabular}}}
\newenvironment{Abstract}{\begin{quotation}  }{\end{quotation}}
\newenvironment{Presented}{\begin{quotation} \begin{center} 
             Presented at the\end{center}
      \begin{center}\begin{large}}{\end{large}\end{center} \end{quotation}}
\def\Acknowledgments{\bigskip  \bigskip \begin{center}
          \large\bf Acknowledgments\end{center}}

\makeatletter
\def\section{\@startsection{section}{0}{\z@}{5.5ex plus .5ex minus
 1.5ex}{2.3ex plus .2ex}{\large\bf}}
\def\subsection{\@startsection{subsection}{1}{\z@}{3.5ex plus .5ex minus
 1.5ex}{1.3ex plus .2ex}{\normalsize\bf}}
\def\subsubsection{\@startsection{subsubsection}{2}{\z@}{-3.5ex plus
-1ex minus  -.2ex}{2.3ex plus .2ex}{\normalsize\sl}}

\renewcommand{\@makecaption}[2]{%
   \vskip 10pt
   \setbox\@tempboxa\hbox{\small #1: #2}
   \ifdim \wd\@tempboxa >\hsize     
       \small #1: #2\par          
     \else                        
       \hbox to\hsize{\hfil\box\@tempboxa\hfil}
   \fi}

 \def\citenum#1{{\def\@cite##1##2{##1}\cite{#1}}}
 
\newcount\@tempcntc
\def\@citex[#1]#2{\if@filesw\immediate\write\@auxout{\string\citation{#2}}\fi
  \@tempcnta\z@\@tempcntb\m@ne\def\@citea{}\@cite{\@for\@citeb:=#2\do
    {\@ifundefined
       {b@\@citeb}{\@citeo\@tempcntb\m@ne\@citea\def\@citea{,}{\bf ?}\@warning
       {Citation `\@citeb' on page \thepage \space undefined}}%
    {\setbox\z@\hbox{\global\@tempcntc0\csname b@\@citeb\endcsname\relax}%
     \ifnum\@tempcntc=\z@ \@citeo\@tempcntb\m@ne
       \@citea\def\@citea{,}\hbox{\csname b@\@citeb\endcsname}%
     \else
      \advance\@tempcntb\@ne
      \ifnum\@tempcntb=\@tempcntc
      \else\advance\@tempcntb\m@ne\@citeo
      \@tempcnta\@tempcntc\@tempcntb\@tempcntc\fi\fi}}\@citeo}{#1}}
\def\@citeo{\ifnum\@tempcnta>\@tempcntb\else\@citea\def\@citea{,}%
  \ifnum\@tempcnta=\@tempcntb\the\@tempcnta\else
  {\advance\@tempcnta\@ne\ifnum\@tempcnta=\@tempcntb \else\def\@citea{--}\fi
    \advance\@tempcnta\m@ne\the\@tempcnta\@citea\the\@tempcntb}\fi\fi}

\def\lsim{\mathrel{\raise.3ex\hbox{$<$\kern-.75em\lower1ex\hbox{$\sim$}}}}
\def\gsim{\mathrel{\raise.3ex\hbox{$>$\kern-.75em\lower1ex\hbox{$\sim$}}}}
\def\Li2{{\rm Li}_2}

\newcommand{\bmr}{\mbox{\boldmath $r$}}

\makeatother

\def\beq{\begin{equation}}
\def\eeq#1{\label{#1}\end{equation}}
\def\eeqn{\end{equation}}

\newenvironment{Eqnarray}%
   {\arraycolsep 0.14em\begin{eqnarray}}{\end{eqnarray}}
\def\beqa{\begin{Eqnarray}}
\def\eeqa#1{\label{#1}\end{Eqnarray}}
\def\eeqan{\end{Eqnarray}}

\let\bar=\overbar

\def\Dslash{\not{\hbox{\kern-4pt $D$}}}
\def\dslash{\not{\hbox{\kern-2pt $\del$}}}

\def\msb{{\bar{\ssstyle M \kern -1pt S}}}

\def\lsim{\mathrel{\raise.3ex\hbox{$<$\kern-.75em\lower1ex\hbox{$\sim$}}}}
\def\gsim{\mathrel{\raise.3ex\hbox{$>$\kern-.75em\lower1ex\hbox{$\sim$}}}}

\begin{document}
\begin{titlepage}
\pubblock

\vfill
\def\thefootnote{\fnsymbol{footnote}}
\Title{Light Quark Mass Effects in \\[5pt] Bottom Quark Mass Determinations}
\vfill
\Author{Andr\'e Hoang\support}
\Address{\csumb}
\vfill
\begin{Abstract}\noindent
Recent results for charm quark mass effects in perturbative bottom
quark mass determinations from $\Upsilon$ mesons are reviewed.
The connection between the behavior of light quark mass
corrections and the infrared sensitivity of some bottom quark mass 
definitions is examined in some detail. 
\end{Abstract}
\vfill
\begin{Presented}
5th International Symposium on Radiative Corrections \\ 
(RADCOR--2000) \\[4pt]
Carmel CA, USA, 11--15 September, 2000
\end{Presented}
\vfill
\end{titlepage}
\def\thefootnote{\arabic{footnote}}
\setcounter{footnote}{0}

\section{Introduction}

The mass of the bottom quark is an important parameter in the
theoretical description of B meson decays. In particular, for the
extraction of $V_{ub}$ and $V_{cb}$ from inclusive decays a precision
in the bottom quark mass of the order 1\% is desirable due to a
strong mass dependence. Precise determinations of
the bottom mass are available from Lattice QCD using the B mesons mass
(see Refs.~\cite{Latticebmass} for recent reviews) and from
perturbative QCD. As the B meson binding energy is dominated by
non-perturbative QCD, perturbative methods can only be applied to
$\Upsilon$ mesons, where the relevant dynamical scales, momentum 
$\langle p\rangle\sim M_b v$ and
energy $\langle E\rangle\sim M_b v^2$, $v$ being the bottom quark
velocity, can be larger than the hadronization scale $\Lambda_{\rm
QCD}$.\footnote{ 
At LEP the $\overline{\mbox{MS}}$ bottom mass at the Z scale has been
determined from the rate of 3 jet events containing a $b\bar b$
pair~\cite{bmassMZ}. This measurement established experimentally the
"running" of the bottom $\overline{\mbox{MS}}$ mass. For this method,
the uncertainties are, however, still too large that light quark mass
effects would be irrelevant.
}

In recent time new perturbative bottom mass analyses have been
carried out including newly available NNLO (i.e. ${\cal O}(v^2,
\alpha_s v, \alpha_s^2)$)
corrections in the non-relativistic expansion for heavy
quark--antiquark systems based on the concept of effective theories
and employing properly defined short-distance 
quark mass definitions\footnote{
I call a heavy quark mass definition without an ${\cal O}(\Lambda_{\rm
QCD})$ ambiguity a short-distance mass.} 
that are adapted to the
non-relativistic framework~\cite{NRQCDbmassNNLO}. 
It has become practice to      
determine the bottom $\overline{\mbox{MS}}$ mass $\overline
m_b(\overline m_b)$ as the reference mass to compare the
various analyses among each other. The various results, which are
based on experimental data on the $\Upsilon$ mesons, show good
agreement with a central value for $\overline m_b(\overline m_b)$ of
about $4.2$~GeV and an uncertainty ranging from $50$ to $80$~MeV.
It is important to note that the uncertainty is predominantly
theoretical, and (as a necessary consequence) partly
depends on the taste and believes of the respective authors. It is therefore
impossible to interpret the value for the error in a statistical
way. It will be the primary aim of future studies and analyses to
achieve a better understanding of this theoretical uncertainty---not
necessarily in order to reduce it further, but in order to put it on
firmer ground.

One effect that has been neglected in previous bottom quark mass
analyses is coming from the finite masses of the light quarks
(u, d, s, c), where "light" means "lighter than the bottom quark
mass". In this talk I report on results and examinations on 
light quark mass corrections at NNLO in the non-relativistic
expansion which have been presented in
Refs.~\cite{Hoang1,Hoang2}. Light quark 
mass corrections are interesting   
because the non-relativistic $b\bar b$ system is governed by a tower
of scales: $M_b$, $\langle p\rangle\sim M_b v$ and 
$\langle E\rangle\sim M_b v^2$. For small velocities these scales form
a hierarchy and their relations to $\Lambda_{\rm QCD}$ determines the
theoretical approach that has to be used to describe the $b\bar b$
dynamics. In the work discussed in this talk I assume that all
three scales are much larger than $\Lambda_{\rm QCD}$ (i.e. that $v$ is not 
smaller than about $0.3$ for bottom quarks) as only for this case the
relevant approach is well understood theoretically.    

Whether the mass of a light quark can also be considered "light" in
the context of the non-relativistic $b\bar b$ dynamics depends on its
relation to the three scales mentioned above.  
The masses of the up, down and strange quarks are indeed much
smaller than any of the three scales and one can expect that for them
the massless approximation is a very good one, as an
expansion in their masses is justified. The mass of the charm,
however, can be about as large as $\langle p\rangle$ and larger
than $\langle E\rangle$, and it is clearly inappropriate to consider
the massless approximation in the first place. I will show that the
effects of the non-zero charm quark mass in bottom mass determinations
can amount to up to several tens of MeV depending of the method and 
the bottom short-distance mass definition that is employed.

Apart from the resulting quantitative effects in the determination of
bottom quark short-distance masses, light quark mass effects are also
interesting conceptually because the massive quark loops provide an
infrared cutoff 
of the momentum flow through gluon lines. The behavior of
light quark mass corrections, in general, can therefore serve as a
natural tool to monitor the degree of infrared (IR) sensitivity of 
various bottom quark mass definition and their resulting
ambiguity. This is in close analogy to the well-known IR renormalon
studies with a fictitious gluon mass, but with the difference
that the light quark mass corrections are real.  

\section{Light Quark Mass Corrections in the Coulomb Potential}

In order to account for the light quark mass effects in the
non-relativistic quark-antiquark dynamics at NNLO we have to determine
the light quark mass corrections to the Coulomb potential 
$V_c(\bmr)=-\frac{C_F \alpha_s}{r}+\ldots$ that occurs in the
Schr\"odinger equation
\begin{eqnarray}
& &
\bigg(\,
-\frac{{\mbox{\boldmath $\nabla$}}^2}
{M_b}
- \frac{{\mbox{\boldmath $\nabla$}}^4}
{4\,M_b^3} 
+ \bigg[\,
  V_{c}}({\bmr) + \ldots
\,\bigg]  
- E
\,\bigg)\,
G({\mbox{\boldmath $r$}},{\mbox{\boldmath $r$}}^\prime, E)
\,  = \, \delta^{(3)}({\mbox{\boldmath $r$}}-{\mbox{\boldmath $r$}}^\prime) 
\,.
\label{NNLOSchroedinger}
\end{eqnarray}
Here $M_b$ is (just as a matter of convenience in writing down
Eq.~(\ref{NNLOSchroedinger})) the bottom pole
mass and $E=E_{cm}-2M_b$, $E_{cm}$ being the center-of-mass energy.
At NNLO, corrections to the Coulomb potential have to be taken into
account up to order $\alpha_s^3$. For massless light quarks these
two-loop corrections have been determined in Refs.~\cite{QCDpotential}. 
At NNLO there is also a potential of order $\alpha_s^2/(M_b r^2)$ 
and another of order $\alpha_s/(M_b^2 r)$. (The latter contains
e.g. the Darwin and the spin-orbit 
interactions.) For these potentials light quark mass corrections
do not have to be considered because they contribute only at NNLO,
whereas for light quark mass corrections we gain at least one more
power of $\alpha_s$. There are also no light quark mass
corrections to the kinetic energy terms. 

\begin{figure}[t]
\begin{center}
\leavevmode
\epsfxsize=4.5cm
\epsffile[220 340 420 480]{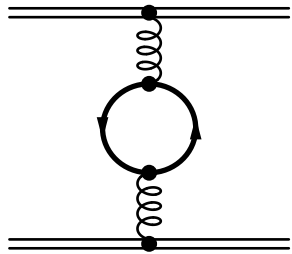}
\vskip  0.5cm
\caption[0]{\label{figVcNLOmassive}
NLO contribution to the static potential coming from the insertion of
a one-loop vacuum polarization of a light quark with finite mass.}
\end{center}
\end{figure}

The light quark mass corrections to the Coulomb potential arise for
the first time at order $\alpha_s^2$   
from the insertion of the light quark self energy into the gluon line
(Fig.~\ref{figVcNLOmassive}). 
%
%
These corrections contribute at NLO in
the non-relativistic power counting. For one massive light quark
flavor, and the other $n_l-1$ light quark flavors being massless,
the correction reads
($\tilde m_q=e^{\gamma_{\mbox{\tiny E}}}\,m_q$):
\begin{eqnarray}
\delta V_{\mbox{\tiny c,m}}^{\mbox{\tiny NLO}}({\mbox{\boldmath
$r$}})
& = &
-\frac{C_F\alpha_s^{(n_l)}}{r}\,
\Big(\frac{\alpha_s^{(n_l)}}{3\,\pi}\Big)\,\bigg\{\,
\ln(\tilde m_q\,r)+\frac{5}{6} +
\int\limits_1^\infty dx
f(x)\,e^{-2 m_q r x}
\,\bigg\}
\,,
\nonumber\\[2mm]
f(x) & \equiv & \frac{1}{x^2}\,
\sqrt{x^2-1}\,\Big(\,1+\frac{1}{2\,x^2}\,\Big)
\,.
\label{VcNLOmassiverspace}
\end{eqnarray}  
The light quark mass corrections vanish for $m_q r\to 0$. This is
related to the definition of $\alpha_s$ that is (here and for the rest
of this presentation) chosen to include the
evolution originating from the massive light quark. In other words,
the massive light quark is not integrated out.
I emphasize that the statements about the size and behavior of the
light quark mass corrections I will make later, 
are only true for this definition of $\alpha_s$. 
It is straightforward to generalize Eq.~(\ref{VcNLOmassiverspace}) to
arbitrary numbers of massive light quark species and different
definitions of $\alpha_s$. However, I emphasize that all previous
analyses, where the light quark masses were neglected have naturally
adopted the same 
scheme, so the results discussed here can be directly interpreted as
additive corrections. 

In Eq.~(\ref{VcNLOmassiverspace}) the NLO light quark mass 
corrections to the Coulomb potential are given in terms of a
subtracted dispersion relation, where $f$ is the absorptive part of
the vacuum 
polarization. This representation is advantageous for determining
the effects of the light quark masses to heavy-quark--antiquark bound
state properties in Rayleigh-Schr\"odinger perturbation theory due to
the simplicity of 
the dependence on $\bmr$. The remaining dispersion integration can
then be carried out numerically (or if possible analytically) at the
very end. In particular, for the determination of multiple
potential insertions at higher orders in time-independent perturbation
theory this method is probably the only feasible one.
On the other hand, the dispersion representation makes the choice of 
definition for $\alpha_s$ manifest.  

At NNLO the light quark mass corrections to the Coulomb potential
arise from dressing the one-loop diagram in
Fig.~(\ref{figVcNLOmassive}) with additional gluon, ghost and light
quark lines. The resulting diagrams where calculated numerically by
Melles~\cite{Melles}. Using again the dispersion 
relation representation it is possible to rewrite Melles' results in a
form that is simple to use for determining the light quark mass
corrections. With the same conventions as for the NLO result,
and adopting the pole mass definition for $m_q$, the result 
reads~\cite{Hoang2}:
\begin{eqnarray}
\lefteqn{
\delta V_{\mbox{\tiny c,m}}^{\mbox{\tiny NNLO}}({\mbox{\boldmath
$r$}})
\, = \,
-\frac{C_F\alpha_s^{(n_l)}}{r}\,
\Big(\frac{\alpha_s^{(n_l)}}{3\,\pi}\Big)^2\,\bigg\{\,
\,
\bigg[\,
3\,\bigg(\ln(\tilde m_q\,r)+\frac{5}{6}\,\bigg)\,
 \bigg(\beta_0\,\ln(\tilde \mu\,r)+\frac{a_1}{2}\bigg) 
+\beta_0\,\frac{\pi^2}{4}
}
\nonumber 
\\[2mm] & & \hspace{0.3cm}
-\frac{3}{2}\,\int\limits_1^\infty dx\,f(x)\,e^{-2 m_q r x}\,
\bigg(\beta_0\,\Big(
\ln\frac{m_q^2}{\mu^2} + g_1(x,m_q,r)
\Big) - a_1
\bigg)
\,\bigg]
\nonumber 
\\[2mm] & & 
-\bigg[\,
-\bigg(\ln(\tilde m\,r)+\frac{5}{6}\,\bigg)^2 - \frac{\pi^2}{12}
\,+\,
\int\limits_1^\infty dx\,f(x)\,e^{-2 m_q r x}\,
\bigg( 
g_2(x) +g_1(x,m_q,r)
- \frac{5}{3}
\bigg)
\,\bigg]
\nonumber 
\\[2mm] & & 
+\bigg[\,\frac{57}{4}\,\bigg(\,
\ln(\tilde m_q\,r) + \frac{161}{228} + \frac{13}{19}\,\zeta_3
+c_1\,\int\limits_{c_2}^\infty \frac{dx}{x}\,e^{-2 m_q r x}    
+ d_1\,\int\limits_{d_2}^\infty \frac{dx}{x}\,e^{-2 m_q r x}
\,\bigg)
\,\bigg]
\,\bigg\}
\,,
\label{VcNNLOmassiverspace}
\end{eqnarray}
where
\begin{eqnarray}
g_1(x,m_q,r) & = &
\ln(4x^2)-{\rm Ei}(2\,m_q\,x\,r)-{\rm Ei}(-2\,m_q\,x\,r)
\,,
\nonumber\\[2mm]
g_2(x) & = &
\frac{5}{3}+\frac{1}{x^2}\bigg(
1+\frac{1}{2\,x}\,\sqrt{x^2-1}\,(1+2x^2)\,
\ln\Big(\frac{x-\sqrt{x^2-1}}{x+\sqrt{x^2-1}}\Big)\bigg)
\,.
\end{eqnarray}
The first three lines in Eq.~(\ref{VcNNLOmassiverspace}) are exact
and involve corrections coming from one and two insertions of the
one-loop massive light quark vacuum polarization. The fourth
line involves all other corrections and is parametrized by four
numerical constants. Of these constants only two are independent,
because the corrections have to vanish for $m_q\to 0$. A useful
parameterization of the constants reads~\cite{Hoang2}
\begin{eqnarray}
c_1 \, = \, \frac{\,\ln\frac{A}{d_2}\,}{\ln\frac{c_2}{d_2}}
\,,\qquad
d_1 \, = \, \frac{\,\ln\frac{c_2}{A}\,}{\ln\frac{c_2}{d_2}}
\,,\qquad
A \, = \, \exp\bigg(
\,\frac{161}{228}+\frac{13}{19}\,\zeta_3-\ln 2
\bigg)
\,,
\end{eqnarray}
where for the constants $c_2$ and $d_2$ one can obtain the following
numerical results from the results of Refs.~\cite{Melles},
\begin{eqnarray}
c_2 \, = \, 0.470 \pm 0.005
\,\qquad
d_2 \, = \, 1.120 \pm 0.010
\,.
\end{eqnarray}

\section{1S Mass, $n=1$ $^3{}S_1$ Binding Energy and Upsilon Expansion}

Naively one might think, that -- at least up to some non-perturbative
effects -- the mass spectrum for $b\bar b$ mesons could be
obtained from solving the Schr\"odinger
equation~(\ref{NNLOSchroedinger}). 
However, the theoretical formalism from which it has been derived is
becoming unreliable for higher radial excitations ($n\gsim 2$),
because the average momentum $\langle p\rangle\sim
\frac{M_b\alpha_s}{n}$ and the average energy $\langle E\rangle\sim
\frac{M_b\alpha_s^2}{n^2}$ become comparable or even smaller than
$\Lambda_{\rm QCD}$.   

For the ground state ($n=1$), however, the formalism for the
perturbative contributions is likely to work, and
one can determine e.g. the  $\overline{\mbox{MS}}$ value of the
bottom quark mass from comparing the $\Upsilon(1S)$ mass with the 
result for the $n=1$, ${}^3S_1$ $b\bar b$ bound state energy once an
estimate for the  non-perturbative corrections is
made~\cite{Pineda1,Yndurain1}. 
 
On the other hand, one can use the perturbation series for a
$n=1$, ${}^3S_1$ bound state mass obtained from
Eq.~(\ref{NNLOSchroedinger}) as a formal short-distance mass
definition, because it is free of the strong 
linear infrared sensitivity that leads to an ambiguity of
order $\Lambda_{\rm QCD}$ in the the pole mass
definition~\cite{Hoang7,Beneke1}. 
In Ref.~\cite{Hoang3} the so called 1S mass has been defined as one half of
the perturbative $n=1$, ${}^3S_1$ bound state
mass obtained from Eq.~(\ref{NNLOSchroedinger}). It has been
demonstrated that the 1S bottom quark mass leads to a very well
convergent perturbative series for totally inclusive B decay
rates~\cite{Hoang4}. Physically this behavior can be understood from the
fact that the 1S mass is adapted to the situation where heavy quarks 
are very close to their mass shell, a situation that is realized for
heavy-heavy as well as for heavy-light bound state systems. 
Heavy quark mass definitions that have this property are called
"threshold masses"~\cite{Hoang5}.
There are other heavy quark threshold mass definitions in the
literature, such as the PS mass~\cite{Beneke1}
and the kinetic mass~\cite{Bigi1}, which will, however, not be further
discussed in this talk.  

The light quark mass corrections to the Coulomb potential presented in
the previous section can be used to determine the light quark mass
corrections in the $b\bar b$ bound state energies. Here we will, as
indicated earlier, only
consider the binding energy of the $n=1$, ${}^3S_1$ triplet ground
state, but the calculation can be generalized without difficulty to
arbitrary quantum numbers. In Dirac notation the NNLO result for the
light quark mass corrections reads 
\begin{eqnarray}
2\,\Big[\,M_b^{\rm 1S}-M_b\,\Big]_{m} & = & 
\Big[\, M_{b\bar b, 1\,{}^3S_1}-2M_b \,\Big]_{m}
\, = \, 
\langle\,1S\,|\,
\delta V_{\mbox{\tiny c,m}}^{\mbox{\tiny NLO}}
\,|\,1S\,\rangle
\nonumber\\[2mm] 
&+&
 \langle\,1S\,|\,
\delta V_{\mbox{\tiny c,m}}^{\mbox{\tiny NNLO}}
\,|\,1S\,\rangle 
\, + \,
\sum\limits_{i\ne\mbox{\tiny 1S}}\hspace{-0.55cm}\int\,\,\,
\langle\,1S\,|\,
\delta V_{\mbox{\tiny c,m}}^{\mbox{\tiny NLO}}\,
\frac{|\,i\,\rangle\,\langle\,i\,|}{E_{\mbox{\tiny 1S}}-E_i}\,
\delta V_{\mbox{\tiny c,m}}^{\mbox{\tiny NLO}}\,
\,|\,1S\,\rangle 
\nonumber
\\[2mm] 
&+&
2\,\sum_{i\ne\mbox{\tiny 1S}}\hspace{-0.55cm}\int\,\,\,
\langle\,1S\,|\,
\delta V_{\mbox{\tiny c,m}}^{\mbox{\tiny NLO}}\,
\frac{|\,i\,\rangle\,\langle\,i\,|}{E_{\mbox{\tiny 1S}}-E_i}\,
V_{\mbox{\tiny c,massless}}^{\mbox{\tiny NLO}}\,
\,|\,1S\,\rangle 
\,.
\label{M1Sformal}
\end{eqnarray}
The first term on the RHS of Eq.~(\ref{M1Sformal}) is the NLO
correction of order $M_b\alpha_s^3$ (in the non-relativistic power
counting) and the terms in the second and third line are the NNLO
corrections of order $M_b\alpha_s^4$. The explicit result based on the
dispersion relation representation is given in Ref~\cite{Hoang2}.
The NLO contribution has already been calculated earlier in
Ref.~\cite{Eiras1} and agrees with the result here.
The corrections coming from gluons and massless quarks 
(not displayed in Eq.~(\ref{M1Sformal}), and called "massless
corrections" from now on) have first been calculated 
in Ref.~\cite{Pineda1}.  

It is not possible to use Eq.~(\ref{M1Sformal}) directly for
any phenomenological analysis that intends a precision better than
order $\Lambda_{\rm QCD}$, because it is written in terms
of the bottom pole mass. Nevertheless,
it is instructive to have a closer look at the behavior of the ligh
quark mass corrections in the bottom 1S-pole mass relation displayed
in Eq.~(\ref{M1Sformal}). For 
$M_b=4.9$~GeV, 
$\overline m_c(\overline m_c)=1.5$~GeV for the $\overline{\mbox{MS}}$
charm quark mass and
$\alpha_s^{(4)}(\mu=4.7~\mbox{GeV})=0.216$ we obtain
\begin{eqnarray}
M_b^{\rm 1S}
& = & 
\Big\{\,
4.9 \, - \,
\Big[\,
0.051
\,\Big]_{\rm LO}
 - \,
\Big[\,
0.074 \, + \,0.0045_{\mbox{\tiny m}}
\,\Big]_{\rm NLO}
\nonumber\\ & &
\hspace{1cm}
 - \,
\Big[\,
0.099 \, + \, 0.0121_{\mbox{\tiny m}}
\,\Big]_{\rm NNLO}
\,\Big\}\,\mbox{GeV}
\,,
\label{1Spoletest}
\end{eqnarray}
where the charm mass corrections are indicated by the subscript m and
the numbers without subscript are from the massless corrections. 
We stress that because 
$M_b\alpha_s\approx 1.5$~GeV the charm mass correction shown in
Eq.~(\ref{1Spoletest})
cannot be obtained by an expansion in the light quark mass. 
To obtain the numbers shown in Eq.~(\ref{1Spoletest}) it is essential  
that the complete expressions for the corrections given in
Eqs.~(\ref{VcNLOmassiverspace}) and (\ref{VcNNLOmassiverspace}) are
taken into account. 
We see that neither the massless corrections nor the charm mass
corrections are converging. This is a consequence of the fact that I
have used the bottom pole mass as an input parameter and the
practical reason why Eq.~(\ref{M1Sformal}) has only limited use for a
phenomenological 
analysis. Another interesting point I would like to mention is that
the NNLO charm mass corrections arising from double
insertions of NLO potentials (the last two terms on the RHS of
Eq.~(\ref{M1Sformal})) make for less than 10\% of the full NNLO
charm mass corrections for $\mu$ between 1.5 and 5~GeV. This property
will be important for the analysis of the charm mass effects in the
$\Upsilon$ sum rules (Sec.~\ref{sectionsumrules}).

It is interesting that the linear
sensitivity to small momenta contained in the pole mass definition is
directly reflected in the analytic behavior of the light quark mass
corrections in Eq.~(\ref{M1Sformal})
for $m_q\to 0$ ($a_s\equiv\alpha^{(n_l)}(\mu)$): 
\begin{eqnarray}
\Big[M_b^{\rm 1S}-M_b\Big]_{m} 
& \longrightarrow &
-\,C_F\,\Big(\frac{a_s}{\pi}\Big)^2\,\bigg\{\,
\frac{\pi^2}{8}\,m_q\,
+\ldots\,\bigg\}_{\rm NLO}
\nonumber
\\[2mm] & &
-\,C_F\,\Big(\frac{a_s}{\pi}\Big)^3\,\bigg\{\,
\frac{\pi^2}{16}\,m_q\,
\bigg[\,
\beta_0\,\bigg(\ln\frac{\mu^2}{m_q^2}-4\ln 2+\frac{14}{3}
\bigg)
\nonumber
\\[2mm] & & \quad
-\,\frac{4}{3}\,\bigg(\frac{59}{15}+2\ln 2\bigg)
+\frac{76}{3\pi}\,\bigg(c_1\,c_2+d_1\,d_2\bigg)
\,\bigg]
+\ldots\,\bigg\}_{\rm NNLO}
\,.
\label{M1Smtozero}
\end{eqnarray}
Using the fact that the dispersion integrations in
Eqs.~(\ref{VcNLOmassiverspace}) and (\ref{VcNNLOmassiverspace}) can be
interpreted as an integration over a gluon mass, one can show (see
Ref.~\cite{Hoang6}) that the linear light quark mass terms in
Eq.~(\ref{M1Smtozero}) are directly related to the linear gluon
mass terms frequently used in renormalon analyses.
A different way to look at this feature is that the mass of the light
quark provides an infrared cutoff for gluon lines due to decoupling at
very small momentum transfers. On the other hand,
if we express the 1S mass in terms of another short-distance mass,
such as the $\overline{\mbox{MS}}$
mass, these linear light quark mass terms do not arise. 
I will come back to this point later in this talk.

I also would like to point out that the structure of the linear light
quark mass corrections in Eq.~(\ref{M1Smtozero}) themselves reveals
their infrared origin. 
First of all, the linear dependence on the light quark mass is
non-analytic since it comes from the square root of $m_q^2$. 
On the other hand, the full vacuum polarization of the
quark loop does only depend on the square of the quark mass,
so the linear mass terms cannot be obtained
from expanding in the light quark mass before doing the integration
over the gluon momentum. Consequently the linear mass terms arise from
gluon momenta of the order of the light quark mass. This feature is
also reflected in the BLM scale in the NNLO term in
Eq.~(\ref{M1Smtozero}) which is of order 
$m_q$ rather than $M_b\alpha_s$. Another observation is that the
non-analyticity of the linear mass terms is associated with an
enhancement factor $\pi^2$. This is a common feature of contributions
that originate from infrared momenta. 
We will see later that this enhancement will help us a lot to
determine the three-loop light quark corrections to the
$\overline{\mbox{MS}}$-pole mass relation. 

Another feature of expression~(\ref{M1Smtozero}) is that 
the linear light quark mass corrections at NLO (NNLO) are multiplied
by $\alpha_s^2$ ($\alpha_s^3$), which appears to contradict the
non-relativistic power counting mentioned before. However, one has to 
take into account that the light quark mass corrections are $M_b$
times a function of the ratio $m_q/(M_b\alpha_s)$. Thus there is no
contradiction with the non-relativistic power counting.\footnote{At
NNLO the energy scale $M_b\alpha_s^2$ does not yet arise as a relevant
dynamical scale.}
On the other hand, one can also show that the linear
light quark mass terms are in fact completely independent of the
quantum number of the bound state. This can be seen from the fact that
the terms displayed in Eq.~(\ref{M1Smtozero}) are equal to the linear
light quark mass terms contained in
$
\frac{1}{2}[
\delta V_{\mbox{\tiny c,m}}^{\mbox{\tiny NLO}}({\mbox{\boldmath
$r$}}) +
\delta V_{\mbox{\tiny c,m}}^{\mbox{\tiny NNLO}}({\mbox{\boldmath
$r$}})]
$.
Because the linear terms are $r$-independent they are just
multiplied by the norm of the bound state wave function, i.e. by 1.
[The suppression of the NNLO charm mass corrections from double
insertions of NLO potentials mentioned above, can be understood from
the dominance of the linear and $r$-independent light quark mass terms
in the potential: constant corrections to potentials give zero
in higher order time-independent perturbation theory.]  
This means that the term $\propto\alpha_s^2$ ($\alpha_s^3$) in
Eq.~(\ref{M1Smtozero}), can equally well be considered as two (three)
loop contributions. This interesting feature is giving us a direct
hint how one has to combine a usual loop expansion in powers of
$\alpha_s$ (such as the perturbative series for the
$\overline{\mbox{MS}}$-pole mass relation) with a non-relativistic
expansion that is in $\alpha_s$ and 
the velocity $v$ (such as the 1S-pole mass relation). The guiding
principle to combine the two types of 
expansions is the cancellation of corrections that are linearly
sensitive to small momenta. The resulting prescription is called
{\it upsilon expansion}~\cite{Hoang4} and has been devised
first for the massless corrections based on more general
arguments. The upsilon expansion states that we have to 
consider corrections of N$^{n-1}$LO in the non-relativistic expansion
(which contains itself a resummation of certain corrections to all
orders in $\alpha_s$) as of order $\alpha_s^n$ in the usual expansion
in the number of loops.

\section{Heavy Quark $\overline{\mbox{MS}}$--Pole Mass Relation}

The pole mass parameter is quite convenient in intermediate steps of
calculating the dynamics of a non-relativistic $Q\bar Q$ pair, because
the Schr\"odinger equation takes its standard QED-like form only in the
pole mass scheme (see Eq.~(\ref{NNLOSchroedinger})). However, for
practical applications (were a 
precision better than $\Lambda_{\rm QCD}$ is relevant) the pole mass
needs to be replaced by a short-distance mass parameter. The standard
choice is the $\overline{\mbox{MS}}$ mass definition. 

The upsilon expansion tells us that, if
we want to describe the non-relativistic $Q\bar Q$ dynamics
at NNLO, the heavy quark $\overline{\mbox{MS}}$-pole mass relation
has to be known at order $\alpha_s^3$. The massless two-loop
corrections have been determined a long time ago in Ref.~\cite{Gray1}
and the massless three-loop corrections can be found in
Refs.~\cite{Melnikov1}. In 
Ref.~\cite{Gray1} also the two-loop light quark mass corrections were
determined fully analytically for any value of the mass. 
For $m_q\ll M_b$ the light quark mass
corrections at two loops read
\begin{eqnarray}
\Big[\overline M_b(\overline M_b)-M_b\Big]_{m} 
& \longrightarrow &
-\,C_F\,\Big(\frac{a_s}{\pi}\Big)^2\,\bigg\{\,
\frac{\pi^2}{8}\,m_q\, - \frac{3}{4} \frac{m_q^2}{M_b}
+\ldots\,\bigg\}
\,.
\label{MSbarmtozerotwoloop}
\end{eqnarray}
It is remarkable that the linear two-loop term is equal to the NLO
term displayed in Eq.~(\ref{M1Smtozero}). This is, of course, not an
accident, but related to the universality of the linear light quark
mass correction, already mentioned before. However, the reason for
this is more general: it has been shown in
Refs.~\cite{Hoang7,Beneke1} that the 
total static energy of a $Q\bar Q$ pair, $E_{\rm stat} = 2
M_Q+V(\bmr)$ is free of any linear dependence on small momenta to all
orders of perturbation theory. Therefore, also the
three-loop linear light quark mass corrections in 
$[\overline M_b(\overline M_b)-M_b]$ are given by the NNLO linear
terms displayed in Eq.~(\ref{M1Smtozero}). From this fact alone
we would not gain much because we cannot get any information on the
size of the three-loop correction with higher powers of the light
quark mass from this argument. However, we have seen before that
the non-analytic 
linear light quark mass terms are enhanced by a factor $\pi^2$ with
respect to the analytic terms with higher powers of the light quark
mass. This feature is also clearly visible in
Eq.~(\ref{MSbarmtozerotwoloop}).  Comparing the size of the full
two-loop light quark mass corrections to the size of the linear term
we find that the difference is at most 15\% for $m/M_b < 0.3$.  
We therefore conclude (or conjecture) that the three-loop linear light
quark mass corrections dominate the yet uncalculated full three-loop
light quark mass corrections in a similar way and that the linear
light quark mass terms provide a very good approximation at the
level of 10\% to the full light quark mass corrections. 
Some more numerical examinations based on BLM type
three loop corrections have been carried out in Ref.~\cite{Hoang2} and
are compatible with the conjecture. It even seems likely that the 
two- and three-loop differences between linear mass approximation and
full results have a different sign, so that the difference in the sum
might be much below 10\%.

To conclude the discussion of the light quark mass corrections to the
$\overline{\mbox{MS}}$--pole mass relation let me also show some
numerical values.  For 
$\overline M_b(\overline M_b)=4.2$~GeV,
$\overline m_c(\overline m_c)=1.5$~GeV for the $\overline{\mbox{MS}}$
charm quark mass and
$\alpha_s^{(4)}(\mu=4.7~\mbox{GeV})=0.216$ we obtain
\begin{eqnarray}
M_b & = &
\Big\{\,
4.2 \, + \,
\Big[\,
0.385
\,\Big]^{{\cal O}(\alpha_s)} + \,
\Big[\,
0.197 \, + \,0.0117_{\mbox{\tiny m}}
\,\Big]^{{\cal O}(\alpha_s^2)}
\nonumber\\ & &
\hspace{1cm} + \,
\Big[\,
0.142 \, + \, 0.0176_{\mbox{\tiny m}}
\,\Big]^{{\cal O}(\alpha_s^3)}
\,\Big\}\,\mbox{GeV}
\,.
\label{msbarpoletest}
\end{eqnarray}
The massless corrections are somewhat better behaved than for the
1S-pole mass relation (but not what one would honestly call
"convergent"), but 
the charm mass corrections (which are in the linear approximation) are
again quite badly behaved.  If we evaluate the charm mass corrections
for $\mu=\overline m_c(\overline m_c)=1.5$~GeV, which is the natural
scale for the linear mass terms, we obtain 36~MeV for the order
$\alpha_s^2$ corrections and 32~MeV for the order 
$\alpha_s^3$ terms.  

\section{Heavy Quark $\overline{\mbox{MS}}$--1S Mass Relation}

The perturbative relation between the bottom 1S and the
$\overline{\mbox{MS}}$ mass can be used for two purposes.
First, one can extract the bottom $\overline{\mbox{MS}}$ mass from the
experimental number for the mass of the $\Upsilon(1S)$ (with a
model-dependence from the estimate of non-perturbative corrections).
Second, one can determine the bottom $\overline{\mbox{MS}}$ mass from
determinations of the 1S mass from methods that are less sensitive to
non-perturbative effects, such as the $\Upsilon$ sum rules. 

Using the results of the two previous sections and combining them
using the upsilon expansion to eliminate the pole mass parameter
(which is absolutely crucial!) it is straightforward to 
derive this relation to order $\alpha_s^3$ (or NNLO in the
non-relativistic expansion). The full analytic expression for the
resulting perturbative series can be found in Ref.~\cite{Hoang2}.

To illustrate the behavior of the series let me show here again some 
numerical results. For 
$M_b^{\rm 1S}=4.7$~GeV, 
$\overline m_c(\overline m_c)=1.5$~GeV and
$\alpha_s^{(4)}(\mu=4.7~\mbox{GeV})=0.216$ we obtain
\begin{eqnarray}
\overline M_b(\overline M_b) & = &
\Big\{\,
4.7 \, - \,
\Big[\,
0.382
\,\Big]^{{\cal O}(\alpha_s), \rm LO} - \,
\Big[\,
0.098 \, + \,0.0072_{\mbox{\tiny m}}
\,\Big]^{{\cal O}(\alpha_s^2), \rm NLO}
\nonumber\\ & &
\hspace{1cm} - \,
\Big[\,
0.030 \, + \, 0.0049_{\mbox{\tiny m}}
\,\Big]^{{\cal O}(\alpha_s^3), \rm NNLO}
\,\Big\}\,\mbox{GeV}
\,.
\label{msbar1Sfirstnumbers}
\end{eqnarray} 
Comparing this result to Eqs.~(\ref{1Spoletest}) and
(\ref{msbarpoletest}) we see that now the massless corrections show a
quite good convergence and the charm mass corrections a fairly
good one. This behavior reflects the fact that the
$\overline{\mbox{MS}}$ and the 1S mass definition both are short-distance
masses, i.e. they do not have an ambiguity of order $\Lambda_{\rm QCD}$
such as the pole mass. It is therefore possible to reliably extract
e.g. the bottom $\overline{\mbox{MS}}$ mass from a given value for the
1S mass with a precision better than $\Lambda_{\rm QCD}$.
 
The reason why the convergence of the charm mass corrections
in our numerical example is not much better is the fact that the
natural choice of the renormalization scale for the charm mass
corrections is of the order of the charm mass and not the bottom mass,
as used in Eq.~(\ref{msbar1Sfirstnumbers}). For $\mu=1.5$~GeV we find 
that the order $\alpha_s^2$ ($\alpha_s^3$) charm mass corrections
amount to $-16$~MeV ($-1$~MeV). However, for $\mu=1.5$~GeV we also
find $(-570,33,55)$~MeV for the order $(\alpha_s, \alpha_s^2,
\alpha_s^3)$ massless corrections, because for them the characteristic 
scale is larger than the charm mass. Because the massless corrections
are the dominant ones it is, of course, more suitable to choose the
larger scale as we did in Eq.~(\ref{msbar1Sfirstnumbers}). 
Thus we find that the charm mass corrections lead to a shift of about
$-15$~MeV in the value of $\overline M_b(\overline M_b)$ for a given
value for $M_b^{\rm 1S}$. 
We note that the size of the charm mass corrections is larger than one
would estimate from an effect of order
$(\frac{\alpha_s}{\pi})^2\frac{m^2}{M_b}$.
The size arises from the
incomplete cancellation of the linear light quark mass term  
in the bottom $\overline{\mbox{MS}}$-pole mass relation, since we
are not allowed to expand in the charm mass in the bottom 1S-pole mass 
relation. 
On the other hand, for the up, down and strange quarks we are allowed
to use the light quark mass expansion (because their masses are
smaller than $\langle p\rangle\sim M_b\alpha_s$ and $\langle
E\rangle\sim M_b\alpha_s^2$) and the linear mass terms are canceled,
leaving a tiny correction that is quadratic in the light quark
masses. For $\overline m_q(\overline m_q)=0.1$~GeV the light quark mass
corrections are well below the 1~MeV level. Thus the mass effects from
the quarks lighter than the charm can be neglected.

The expression for the order $\alpha_s^3$ (NNLO) relation between the
bottom 1S and $\overline{\mbox{MS}}$ is quite complicated, but it
turns out that for $\overline m_q(\overline m_q)>0.4$~GeV and $\mu\gsim
2.5$~GeV the dependence of the bottom $\overline{\mbox{MS}}$-1S mass
relation at order $\alpha_s^3$ on all parameters is approximately
linear. This allows for the derivation of a handy approximation
formula~\cite{Hoang2}, which is applicable to all cases of interest and
allows for a quick determination of the charm quark mass effects:
\begin{eqnarray}
\overline M_{\mbox{\tiny b}}(\overline M_{\mbox{\tiny b}})
& = &
\bigg[\,
4.169\,\mbox{GeV} 
\, - \, 0.01\,\Big(\,
  \overline m_c(\overline m_c)-1.4\,\mbox{GeV}\,\Big)
\, + \, 0.925\,\Big(\,
  M_b^{\rm 1S}
   -4.69\,\mbox{GeV}\,\Big)
\nonumber
\\[2mm] & & \hspace{0.5cm}
\, -\, 9.1\,\Big(\,
  \alpha_s^{(5)}(M_Z) - 0.118
\,\Big)\,\mbox{GeV}
\, + \, 0.0057\,\Big(\,
   \mu - 4.69
\,\Big)\,\mbox{GeV}
\,\bigg]
\,.
\label{msbarapproximation}
\end{eqnarray}

\section{Bottom $\overline{\mbox{MS}}$ Mass from $M(\Upsilon(1S))$}

We can apply Eq.~(\ref{msbarapproximation}) to extract
the bottom  $\overline{\mbox{MS}}$ mass from the mass of the
$\Upsilon(1S)$ meson, if we assume that $\langle p\rangle$ and
$\langle E\rangle$ both are larger than $\Lambda_{\rm QCD}$ for the 1S
state. Because for higher radial excitations $\Upsilon(2S), \ldots$
this assumption is more difficult to justify, we do not attempt
a similar analyses for them. Recalling that the 1S mass just
incorporates the perturbative effects, we need an estimate of the size
of non-perturbative effects in the $\Upsilon(1S)$ bound state. Using
the gluon condensate contribution obtained by Voloshin and
Leutwyler~\cite{VoloshinLeut} 
we get
\begin{eqnarray}
\Big[\,M(\Upsilon(\mbox{1S}))\,\Big]^{\mbox{\tiny non-pert}}
&\approx &
\frac{1872}{1275}\,\frac{M_{\mbox{\tiny b}}\,\pi}
{(M_{\mbox{\tiny b}}\,C_F\,\alpha_s)^4}\,
\langle\,\alpha_s\,
{\mbox{\boldmath
$G$}}^2\,\rangle
\,.
\label{upsilon1Scondensate}
\end{eqnarray}
Using the standard literature range
$\langle\,\alpha_s\,{\mbox{\boldmath $G$}}^2\,\rangle=0.05\pm
0.03\,\mbox{GeV}^4$ the non-perturbative correction can range from
anywhere between $10$ and $200$~MeV due to the strong dependence on
the renormalization scale in $\alpha_s$. 
Taking this estimate and $M(\Upsilon(1S))=9460$~MeV we arrive at
\begin{equation}
M_{\mbox{\tiny b}}^{\mbox{\tiny 1S}}
\, = \, 
\frac{1}{2}\,
\Big\{\,
M(\Upsilon(\mbox{1S})) -
\Big[\,M(\Upsilon(\mbox{1S}))\,\Big]^{\mbox{\tiny non-pert}}
\,\Big\}
\, = \, 
4.68\,\pm\,0.05\,\mbox{GeV}
\,.
\label{M1Sestimate}
\end{equation}
From Eq.~(\ref{msbarapproximation}) we then obtain 
\begin{equation}
\overline M_b(\overline M_b) \, = \,
4.16\,\pm\,0.06\,\mbox{GeV}
\end{equation}
for the bottom $\overline{\mbox{MS}}$ mass 
for $\overline m_c(\overline m_c)$ around $1.3$~GeV and adding the
uncertainty in $\alpha_s(M_Z)$ quadratically. The charm mass
corrections amount to about $-15$~MeV and are smaller than the
uncertainty.

\section{$\Upsilon$ Sum Rules}
\label{sectionsumrules}

A method that is in principle much less sensitive to non-perturbative
effects is to extract the bottom quark mass from moments of the $b\bar
b$ total cross section in $e^+e^-$ annihilation:
\begin{equation}
P_n \, = \,
\int\limits^\infty_{s_{\rm min}} \frac{d s}{s^{n+1}}\,R(s)
\,.
\label{momentsdef2}
\end{equation}
Here $R$ is the inclusive $b\bar b$ cross section normalized to the
muon pair cross section and $s$ the square of the c.m.\ energy. The
idea of the $\Upsilon$ sum rules is to  
determine the bottom quark mass from comparing theoretical
calculations of the moments $P_n$ with moments obtained from
experimental data~\cite{Voloshin1}. 
Non-perturbative effects can, in contrast
to calculations of the $b\bar b$ spectrum, be suppressed by hand by
choosing the parameter $n$ small enough such that the size of the
effective integration range in the c.m.\ energy in~(\ref{momentsdef2})
is much larger than $\Lambda_{\rm QCD}$~\cite{Poggio1}. For 
$n\lsim 15-20$ the gluon condensate 
corrections to the theoretical moments turn out to be smaller than a
percent and can be neglected~\cite{Voloshin1}. On the other hand, one
would like to 
suppress the influence of the quite badly known $b\bar b$ continuum in
the experimental moments. This can be achieved by choosing $n$ large,
so that non-relativistic dynamics dominates the (theoretical and
experimental) moments. In this case the only experimental
input needed for the determination of the experimental moments are the
masses and the electronic partial widths of the $\Upsilon$ mesons. 
The continuum can be approximated by a crude model.
Due to the large size of the bottom quark mass one can easily find a
window, $4\lsim n\lsim 15$, for which both requirements can be met. 
One can show that the average relative velocity of $b\bar b$ pairs
that dominate the moments is of order $v_{\rm eff}=1/\sqrt{n}$. So, by
restricting $n$ to the values just mentioned we find that $\langle
p\rangle\sim M_b v_{\rm eff}$ and $\langle E\rangle\sim M_b v_{\rm
eff}^2$ form a hierarchy and are larger than $\Lambda_{\rm QCD}$.   
Thus the Schr\"odinger equation~(\ref{NNLOSchroedinger}) can be safely
used to describe the dynamics encoded in the moments.

For the case of massless light quarks a number of 
NLO~\cite{Voloshin1,Penin1} and 
NNLO~\cite{NRQCDbmassNNLO,NNLOsumrulesnaive} analyses
have been carried out. For the restricted range of $n$ the
theoretical moments are directly related to the Green function
$G(0,0,E)$ of the Schr\"odinger equation~(\ref{NNLOSchroedinger}). It
is also necessary to include, for the NNLO moments, a two-loop
renormalization of an external current that describes the annihilation
of a $b\bar b$ pair into a photon. One can either calculate the bound
state resonances and the continuum explicitly and carry out the energy
integration in Eq.~(\ref{momentsdef2}) on the real energy
axis, or one uses the analyticity properties of the
Green function and integrates instead in the negative complex energy
plane. The calculations involved in these computations are
quite extensive and shall not be describe here in more detail.

The light quark mass corrections to the moments have been determined
in Ref.~\cite{Hoang2}.
The NLO and NNLO corrections to the Green function 
$G(0,0,E)$ are determined with time-independent perturbation theory in
analogy to Eq.~(\ref{M1Sformal}). In Ref.~\cite{Hoang2} only the light
quark mass corrections at NLO were fully determined, whereas at NNLO
only the single insertion contribution (corresponding to the
second term on the RHS of Eq.~(\ref{M1Sformal})) was calculated.
The NNLO double insertion contributions (last two terms on the RHS of 
Eq.~(\ref{M1Sformal})) were neglected based on the assumption that
the suppression of the double insertion corrections (last two terms on
the RHS of Eq.~(\ref{M1Sformal})) is as effective as for the 
calculation of the 1S mass. 

The light quark mass corrections to the
two-loop renormalization of the current were neglected
because they are expected to be of order $(\alpha_s/\pi)^2 (m/M_b)^2$,
which is at the permille level even for charm quarks.
There are no linear light quark mass corrections to the current
renormalization because it only contains effects from 
momenta of order $M_b$. This means that non-analytic, and in
particular $\pi^2$-enhanced linear light quark mass corrections do not 
exist. 

In Ref.~\cite{Hoang2} a detailed analysis of the light quark mass
correction in the 1S mass scheme has been carried out. The mass
effects from up, down and strange quarks are negligible. For typical
choices for the renormalization scales, $\alpha_s(M_Z)$ and the 1S
mass we find that the NNLO charm mass corrections are around -1\% for
$n=4$ and around -5\% for $n=10$, for $\overline m_c(\overline
m_c)\approx 1.5$~GeV. 
Thus the charm mass corrections in the bottom 1S mass are negative. 
From dimensional analysis we see that the moments $P_n$ are
proportional to $(M_b^{\rm 1S})^{-2n}$, so we can estimate that the
charm mass 
corrections amount to about -15~MeV. 
(In the pole mass scheme the
corrections are considerably larger due to the large non-analytic
charm mass corrections that we have already discussed in the bottom
1S-pole mass relation. By using the bottom 1S mass in the moments
these large corrections are canceled. The same is true for the
massless corrections, see e.g. Ref.~\cite{Hoang8} for a comparison of
results in different mass schemes.)

In Ref.~\cite{Hoang2} I have carried out a more thorough NLO and NNLO
analysis based on a $\chi^2$-procedure, where several (theoretical
and experimental) moments have been fitted simultaneously.
This fitting procedure puts more statistical weight on the relative
than on the absolute size of the moments. Interestingly, the relative
size of the theoretical moments turns out to have smaller perturbative
corrections than their absolute size. This is an issue that
is well known  from the total cross section for $t\bar t$ production
close to threshold in $e^+e^-$ annihilation, where the line-shape for
the NNLO prediction has much smaller perturbative corrections than the
height (see Ref.~\cite{Hoang5} for a review). Recently, using 
renormalization group improved perturbation theory in the framework of
"vNRQCD"~\cite{Luke1} the 
height of the line-shape has been considerably stabilized by summation
of logarithms of the top velocity at NNLL~\cite{Hoang9}. I will
be quite interesting to see whether a summation of these logarithms
can improve the behavior of the absolute size of the moments as well.

%
%
\begin{figure}[t]
\begin{center}
\leavevmode
\epsfxsize=3.5cm
\epsffile[230 580 440 710]{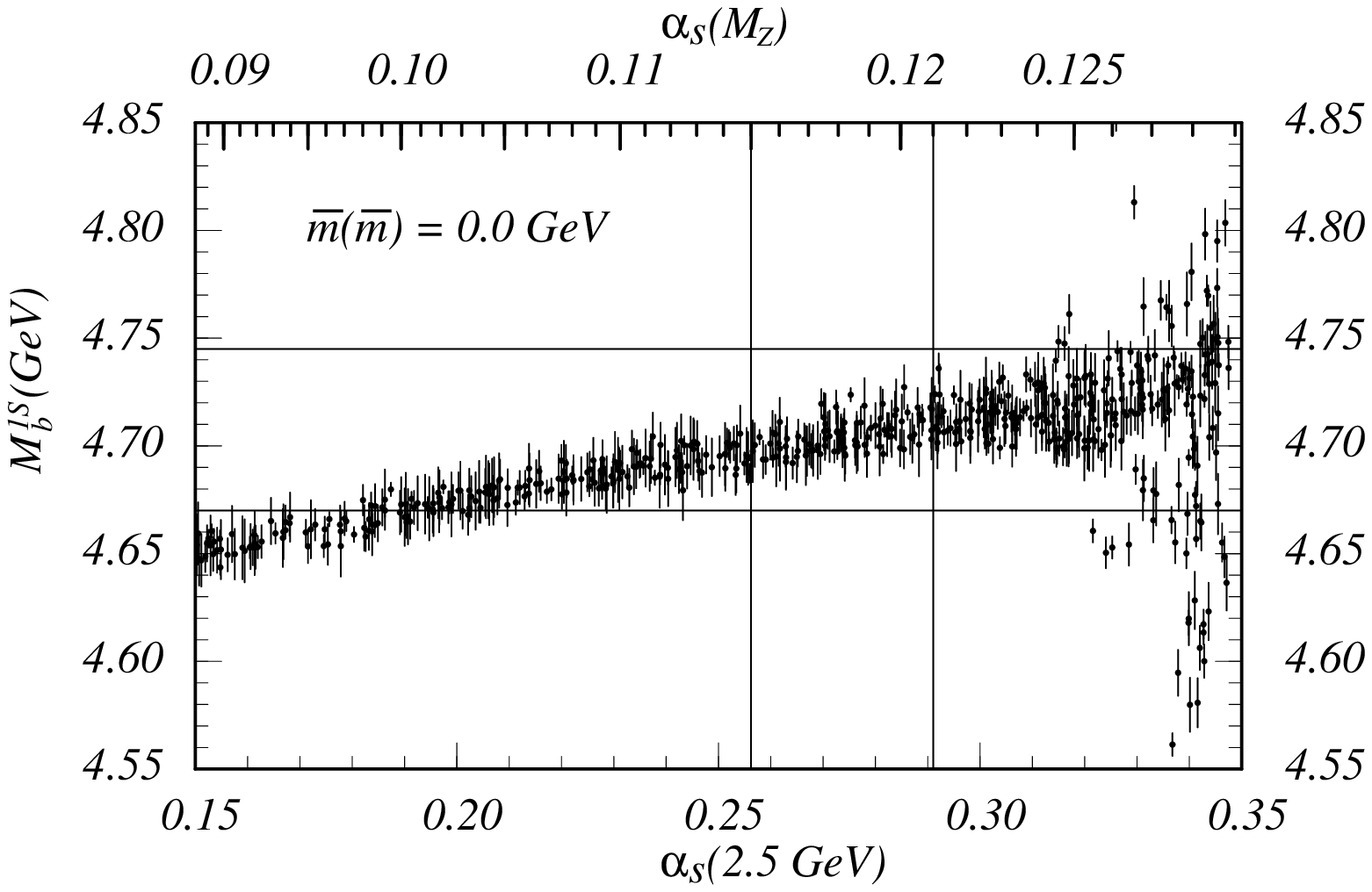}
\hspace{4cm}
\leavevmode
\epsfxsize=3.5cm
\epsffile[230 580 440 710]{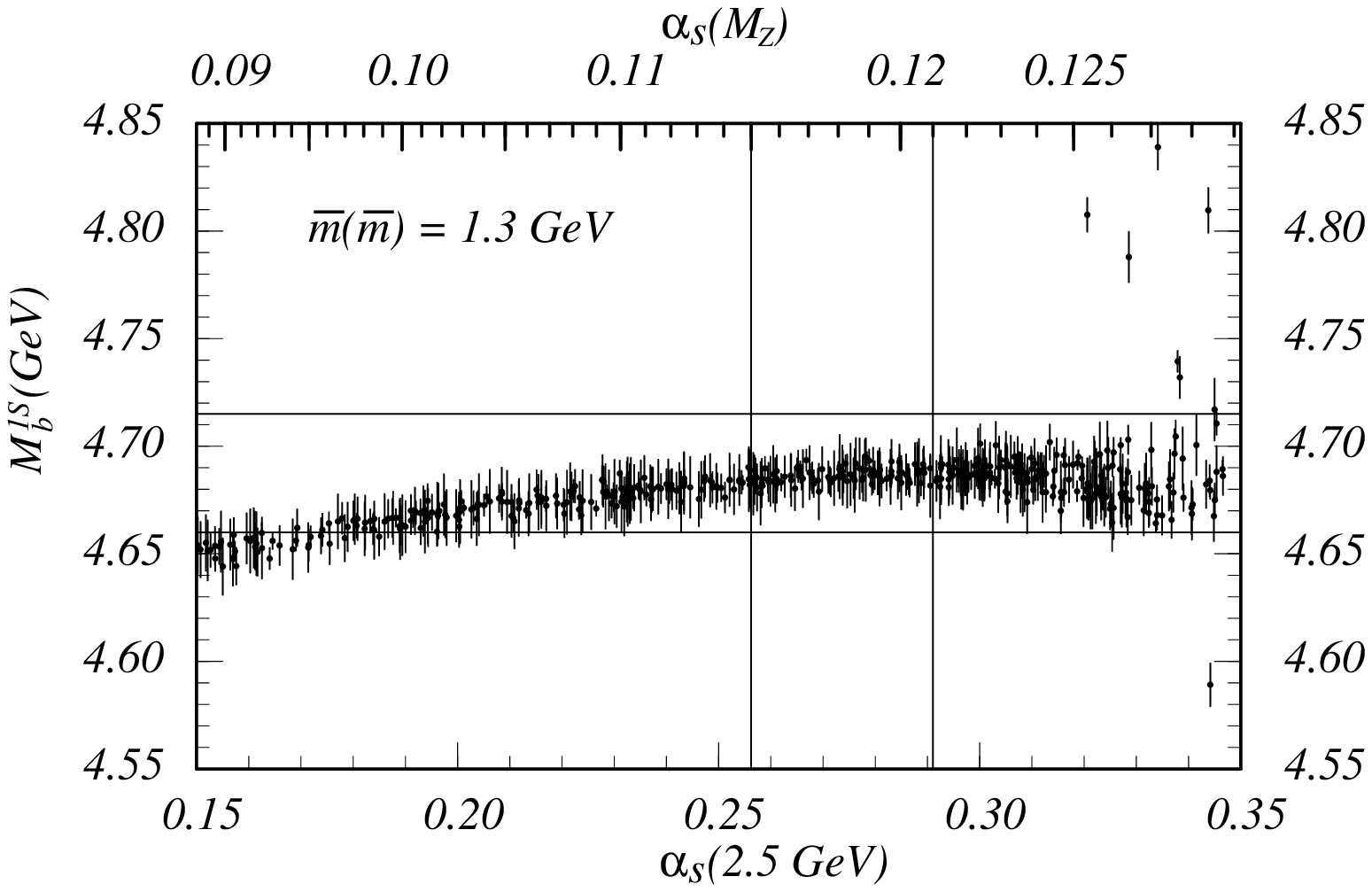}
\vskip  2.3cm
\caption[0]{\label{figsumrule}
Results for the allowed range of $M_b^{\rm 1S}$ for given
values of $\alpha_s^{(5)}(M_Z)$ at NNLO for 
$\overline m_c(\overline m_c)=0.0$ and $1.3$~GeV. 
It is illustrated by the vertical and horizontal lines how the allowed
range for $M_b^{\rm 1S}$ is obtained, if 
$0.115\le\alpha_s^{(5)}(M_Z)\le 0.121$ is taken as an input. 
}
\end{center}
\end{figure}
%
%
The result for the allowed values for the bottom 1S mass from the
$\chi^2$-procedure based on the NNLO theoretical moments 
is displayed in Fig.~\ref{figsumrule} for $\overline
m_c(\overline m_c)=0.0$ and $1.3$~GeV. 
The dots represent points of minimal $\chi^2$ for a large number of
random choices of renormalization scales and sets of $n$'s for given
values of $\alpha_s^{(5)}(M_Z)$. Experimental errors at $95\%$ CL are
displayed as vertical lines. (See Ref.~\cite{Hoang2} for more details.)
The dependence of the 1S mass on the input value for
$\alpha_s^{(5)}(M_Z)$ turns out to be quite weak, particularly if the
charm mass corrections are taken into account.
For $\overline m_c(\overline m_c)=1.4\pm 0.3$~GeV we obtain
\begin{eqnarray}
M_b^{\rm 1S}=4.69 \, \pm \, 0.03\,\,\mbox{GeV}
\label{M1Sfinalresult}
\end{eqnarray}
for the bottom 1S mass. The charm mass corrections shift the 1S mass
by about -20~MeV, which is compatible with the crude estimate
mentioned above. 
It is quite interesting that this result is compatible with the
determination of the 1S mass from $M(\Upsilon(1S))$ (see
Eq.~(\ref{M1Sestimate})). The difference to Eq.~(\ref{M1Sestimate}) is
that that the central value and the error from the sum rule
determination are completely independent of the estimate for the size
of the non-perturbative gluon condensate contribution. 
In this respect the result in Eq.~(\ref{M1Sfinalresult}) should be
considered as more solid than the one in Eq.~(\ref{M1Sestimate}). 
  
Taking the approximation formula~(\ref{msbarapproximation}) we then
arrive at 
\begin{eqnarray}
\overline M_b(\overline M_b)
& = &
4.17 \, \pm \, 0.05\,\mbox{GeV}
\,,
\label{msbarfinalresults}
\end{eqnarray}
for the bottom  $\overline{\mbox{MS}}$ mass, where we have added
quadratically the uncertainties from $M_b^{\rm 1S}$ (30~MeV),
$\alpha_s^{(5)}(M_Z)=0.118\pm 0.003$ (30~MeV), $\mu=4.7\pm 3$GeV
(15~MeV) and $\overline m_c(\overline m_c)=1.4\pm 0.3$~GeV (5~MeV). 
We note that the charm mass corrections in the $\Upsilon$ sum rules
and those in the $\overline{\mbox{MS}}$-1S mass relation are additive
and lead to an overall shift of about -30~MeV in $\overline
M_{\mbox{\tiny b}}(\overline M_{\mbox{\tiny b}})$ compared to an
analysis where the charm mass is neglected altogether.  Thus the charm
mass effects are are relevant for the 
sum rule method, in particular if it should turn out that the
perturbative uncertainties in the moments can be decreased in the
future. 

\section{Conclusions}

In this talk I have reviewed recent results for light quark mass
effects in perturbative bottom quark mass determinations from
$\Upsilon$ mesons~\cite{Hoang1,Hoang2}. 
We find that the effects of the charm mass are non-negligible in view
of the present theoretical uncertainties. Light quark mass corrections
were determined for the 1S-pole mass relation at NNLO, for the 
$\overline{\mbox{MS}}$-pole mass relation at order $\alpha_s^3$ and
for the $\Upsilon$ sum rules at NNLO. In the determination of the bottom
$\overline{\mbox{MS}}$ mass, $\overline M_b(\overline M_b)$, from the
mass of the $\Upsilon(1S)$ meson the charm mass corrections shift the
value of $\overline M_b(\overline M_b)$ by about -15~MeV. In the
determination of the bottom 1S mass, $M_b^{\rm 1S}$, from the
$\Upsilon$ sum rules the charm mass corrections shift the value of
$M_b^{\rm 1S}$ by about -20~MeV. The overall shift of $\overline
M_b(\overline M_b)$ obtained from the $\Upsilon$ sum rule analysis is
about -30~MeV. The sign and size of the charm mass corrections shows
agreement with a recent unquenched lattice determination
of $\overline M_b(\overline M_b)$~\cite{Latticeunquenched}.

On the conceptual side light quark mass corrections are interesting
because they provide a natural tool to investigate the sensitivity the
heavy quark mass definitions to infrared momenta.

\Acknowledgments
This work was supported in part by the EU Fourth Framework Program
``Training and Mobility of Researchers'', Network ``Quantum
Chromodynamics and Deep Structure of Elementary Particles'', contract
FMRX-CT98-0194 (DG12-MIHT).


\begin{thebibliography}{99}


%
\bibitem{Latticebmass}
C.~T.~Sachrajda,
hep-lat/0101003;
V.~Lubicz,
hep-lat/0012003.
%
\bibitem{bmassMZ}
G.~Dissertori,
hep-ex/0010005.
%
A.~Brandenburg, P.~N.~Burrows, D.~Muller, N.~Oishi and P.~Uwer,
Phys.\ Lett.\ B {\bf 468}, 168 (1999).
%
\bibitem{NRQCDbmassNNLO}
K.~Melnikov and A.~Yelkhovsky,
Phys.\ Rev.\ D {\bf 59}, 114009 (1999);
%
A.~H.~Hoang,
Phys.\ Rev.\ D {\bf 61}, 034005 (2000);
%
M.~Beneke and A.~Signer,
Phys.\ Lett.\ B {\bf 471}, 233 (1999).
%
\bibitem{Hoang1}
A.~H.~Hoang and A.~V.~Manohar,
Phys.\ Lett.\ B {\bf 483}, 94 (2000).
%
\bibitem{Hoang2}
A.~H.~Hoang,
hep-ph/0008102.
%
\bibitem{QCDpotential}
Y.~Schroder,
Phys.\ Lett.\ B {\bf 447}, 321 (1999);
%
M.~Peter,
Phys.\ Rev.\ Lett.\ {\bf 78}, 602 (1997).
%
\bibitem{Melles}
M.~Melles,
Phys.\ Rev.\ D {\bf 62}, 074019 (2000);
Phys.\ Rev.\ D {\bf 58}, 114004 (1998).
%
\bibitem{Pineda1}
A.~Pineda and F.~J.~Yndurain,
Phys.\ Rev.\ D {\bf 58}, 094022 (1998);
%
S.~Titard and F.~J.~Yndurain,
Phys.\ Rev.\ D {\bf 49}, 6007 (1994).
%
\bibitem{Yndurain1}
F.~J.~Yndurain,
Nucl.\ Phys.\ Proc.\ Suppl.\ {\bf 93}, 196 (2001).
%
%
%
%
\bibitem{Hoang7}
A.~H.~Hoang, M.~C.~Smith, T.~Stelzer and S.~Willenbrock,
Phys.\ Rev.\ D {\bf 59}, 114014 (1999).
%
\bibitem{Beneke1}
M.~Beneke,
Phys.\ Lett.\ B {\bf 434}, 115 (1998).
%
\bibitem{Hoang3}
A.~H.~Hoang and T.~Teubner,
Phys.\ Rev.\ D {\bf 60}, 114027 (1999).
%
\bibitem{Hoang4}
A.~H.~Hoang, Z.~Ligeti and A.~V.~Manohar,
Phys.\ Rev.\ D {\bf 59}, 074017 (1999);
%
Phys.\ Rev.\ Lett.\ {\bf 82}, 277 (1999).
%
\bibitem{Hoang5}
A.~H.~Hoang {\it et al.},
Eur.\ Phys.\ J.\ directC {\bf 3}, 1 (2000).
%
\bibitem{Bigi1}
I.~Bigi, M.~Shifman, N.~Uraltsev and A.~Vainshtein,
Phys.\ Rev.\ D {\bf 56}, 4017 (1997).
%
\bibitem{Eiras1}
D.~Eiras and J.~Soto,
Phys.\ Lett.\ B {\bf 491}, 101 (2000).
%
\bibitem{Hoang6}
A.~Hoang,
PhD thesis, University of Karlsruhe, RX-1578 (AACHEN).
%
\bibitem{Gray1}
N.~Gray, D.~J.~Broadhurst, W.~Grafe and K.~Schilcher,
Z.\ Phys.\ C {\bf 48}, 673 (1990).
%
\bibitem{Melnikov1}
K.~Melnikov and T.~v.~Ritbergen,
Phys.\ Lett.\  {\bf B482}, 99 (2000).
%
K.~G.~Chetyrkin and M.~Steinhauser,
Nucl.\ Phys.\  {\bf B573}, 617 (2000).
%
\bibitem{VoloshinLeut}
M.~B.~Voloshin, Yad.\ Fiz.\ {\bf 36}, 247 (1982)
[Sov.\ J.\ Nucl.\ Phys.\ {\bf 36},1 (1982)];
%
H.~Leutwyler,
Phys.\ Lett.\  {\bf B98}, 447 (1981).
%
\bibitem{Voloshin1}
M.~B.~Voloshin,
Int.\ J.\ Mod.\ Phys.\  {\bf A10}, 2865 (1995).
%
\bibitem{Poggio1}
E.~C.~Poggio, H.~R.~Quinn and S.~Weinberg,
Phys.\ Rev.\ D {\bf 13}, 1958 (1976).
%
\bibitem{Penin1}
J.~H.~Kuhn, A.~A.~Penin and A.~A.~Pivovarov,
Nucl.\ Phys.\ B {\bf 534}, 356 (1998).
%
\bibitem{NNLOsumrulesnaive}
A.~H.~Hoang,
Phys.\ Rev.\ D {\bf 59}, 014039 (1999);
%
A.~A.~Penin and A.~A.~Pivovarov,
Phys.\ Lett.\ B {\bf 443}, 264 (1998);
%
Nucl.\ Phys.\ B {\bf 549}, 217 (1999);
%
\bibitem{Hoang8}
A.~H.~Hoang,
Nucl.\ Phys.\ Proc.\ Suppl.\ {\bf 86}, 512 (2000).
%
\bibitem{Luke1}
M.~E.~Luke, A.~V.~Manohar and I.~Z.~Rothstein,
Phys.\ Rev.\ D {\bf 61}, 074025 (2000);
A.~V.~Manohar and I.~W.~Stewart,
Phys.\ Rev.\ D {\bf 62}, 014033 (2000);
Phys.\ Rev.\ D {\bf 62}, 074015 (2000);
Phys.\ Rev.\ D {\bf 63}, 054004 (2001).
%
A.~H.~Hoang, A.~V.~Manohar and I.~W.~Stewart,
hep-ph/0102257.
%
\bibitem{Hoang9}
A.~H.~Hoang, A.~V.~Manohar, I.~W.~Stewart and T.~Teubner,
hep-ph/0011254.
%
\bibitem{Latticeunquenched}
V.~Gimenez, L.~Giusti, G.~Martinelli and F.~Rapuano,
JHEP{\bf 0003}, 018 (2000).












\end{thebibliography}
\end{document}